\documentclass[aps,prd,showpacs,preprintnumbers,twocolumn]{revtex4}
\usepackage[dvips]{graphicx}
\usepackage{bm,latexsym,amsmath,amssymb,amsfonts}
\newcommand*{\rml}{{\rm C}}
\newcommand*{\rmr}{{\rm B}}
\begin{document}
\title{UV caps, IR modification of gravity, and recovery of 4D gravity in\\ regularized braneworlds}

\author{Tsutomu~Kobayashi} %
\email[Email: ]{tsutomu"at"gravity.phys.waseda.ac.jp}
\affiliation{Department of Physics, Waseda University, Okubo 3-4-1, Shinjuku, Tokyo 169-8555, Japan}

\begin{abstract}
In the context of six-dimensional conical braneworlds we consider a simple and explicit model
that incorporates long distance modification of gravity and
regularization of codimension-2 singularities.
To resolve the conical singularities we replace the codimension-2 branes with
ring-like codimension-1 branes, filling in the interiors with regular caps.
The six-dimensional Planck scale in the cap is assumed to be much greater
than the bulk Planck scale, which
gives rise to the effect analogous to brane-induced gravity.
Weak gravity on the regularized brane is studied in the case of a sharp conical bulk.
We show by a linear analysis that gravity at short distances is effectively described by
the four-dimensional Brans-Dicke theory, while the higher dimensional nature of gravity emerges
at long distances.
The linear analysis breaks down
at some intermediate scale, below which four-dimensional Einstein gravity is shown to be
recovered thanks to the second-order effects of the brane bending.
\end{abstract}

\preprint{WU-AP/287/08}
\maketitle

\section{Introduction}

The most intriguing possibility for the origin of
the current acceleration of the Universe~\cite{Acceleration}
may be modifying gravity at very long distances.
Within the framework of braneworlds,
Dvali, Gabadadze, and Porrati (DGP) proposed
a model in which gravity looks five-dimensional (5D)
at long distances and 4D at short distances~\cite{DGP}.
In their model,
gravity becomes weaker on large scales due to its leakage from our 4D brane
to the higher dimensional space, while
the induced-gravity term on the brane maintains
the 4D nature of short-range gravity.
Although the original motivation of the idea of~\cite{DGP} was
to realize 4D gravity even in the presence of infinitely large extra dimensions,
it was soon noticed that the model gives rise to the self-acceleration, i.e.,
the accelerating expansion of the Universe without a cosmological constant~\cite{SA, Lue_Rev, Koyama_CC}.
The modified Friedmann equation in the DGP braneworld is given by~\cite{SA}
\begin{eqnarray*}
H^2-\frac{H}{r_c}=\frac{8\pi G}{3}\rho,
\end{eqnarray*}
where $r_c$ is the crossover scale above which gravity is modified.
This allows for the late-time accelerating solution, $H\to H_{\infty}=r_c^{-1}$.
Despite multiple problematic features of the DGP
model~\cite{Rubakov:2003zb, Ghost0, Ghost1, Ghost2,Kaloper:2005az,Gregory:2007xy},
it is still worth exploring
the arena of IR modified gravity in the braneworld context
because more elaborated brane models may provide
consistent IR modification which has possible relevance to the cosmological constant problem,
in terms of a fully covariant and nonlinear theory of gravity.

So far the consequences of brane-induced gravity
have been often discussed in codimension-1 models, but
it seems quite interesting to consider more general, codimension-$N\geq 2$ braneworlds
with induced gravity.
Infinitely thin branes with $N\geq 2$, however, suffer from certain singularities
which must be dealt with some care~\cite{DG}.
The pathology
is clearly illustrated by the shock wave solution on thin codimension-2 branes~\cite{Kaloper}:
the shock wave profile is perfectly 4D over all distance scales on the brane,
but it vanishes (even infinitesimally) away from the brane.
Namely, gravity confinement is too effective.
This is due to the fact that the Green function in the space transverse to the brane
diverges at the origin. (The divergence is the generic feature for $N\geq2$.)
If one is to resolve this short distance singularity somehow,
the regularization at UV will affect the behavior of gravity
at IR~\cite{Kiri, Kaloper, Seesaw, Reg-Rubakov, Reg-DGP, Softly}.
(See also different approaches to construct higher codimension braneworlds with induced
gravity~\cite{Int, Cascading}.)

In this paper we consider a 6D brane model
that incorporates long distance modification of
gravity and regularization of codimension-2 singularities.
The codimension-2 branes in our model are regularized
in the manner of~\cite{Kaloper}, replacing
them with ring-like codimension-1 branes and filling in the interiors with regular caps.
The scheme here is similar to what is widely used in flux-stabilized models of 6D
braneworlds~\cite{Peloso, Papa, Reg-TK, UVcaps}.
Then, instead of introducing brane-induced gravity term directly,
we use the possible mechanism for realizing {\em effectively} DGP-type gravity, that is,
asymmetry between the two sides of the brane~\cite{Padilla, KoyamaKoyama}.
While the standard way to study codimension-1 brane models in five dimensions
is assuming $\mathbb{Z}_2$-symmetry across the brane~\cite{RS, Z2},
the bulk-cap system in regularized braneworlds has an asymmetric configuration in general~\cite{nonZ2}.
In addition to this natural geometrical asymmetry, we allow the cap Planck scale, $M_\rml$,
to differ from that in the bulk, $M_\rmr$.
If $M_\rml\gg M_\rmr$, this leads to an explicit realization of
the regularization of brane-induced gravity~\cite{Reg-DGP}.
The purpose of the present paper is
to investigate in detail the behavior of weak gravity in such a braneworld.
Even if gravity is weak,
linearization is not justified below a certain distance scale, as in the DGP
model~\cite{Deffayet,Lue:2001gc,Gabadadze:2004iy,Gruzinov:2001hp,Tanaka, KScp}.
We identify the scale and then
carefully take into account the second order effects of the brane bending
to explore the seminonlinear regime.

This paper is organized as follows. In the next section
we present our background model of regularized braneworlds.
We then study the behavior of linear perturbations
in Sec.~III, showing that brane gravity
at short distances
is described by the 4D Brans-Dicke theory.
As an example, in Sec.~\ref{sec:pointsource} we compute
the gravitational field of a static point source modeled by a loop of a string along the ring-like brane.
Using this example we confirm that gravity looks five- or six-dimensional at long distances.
In Sec.~IV we perform a nonlinear analysis
to see how 4D Einstein gravity emerges at even shorter distances.
Sec.~V is devoted to conclusions.

\section{The model}

Let us consider a model
in which the bulk is given by 4D Minkowski spacetime $\times$ a infinite cone,\footnote{
We use $A, B, ...$ for 6D indices,
$\mu, \nu, ...$ for 4D ones, and $a, b, ...$ for 5D ones parallel to the 4-brane.
}
\begin{eqnarray}
g_{AB}dx^Adx^B= \eta_{\mu\nu}dx^{\mu}dx^{\nu}+dr^2+\beta^2r^2d\varphi^2,
\label{bulkmetric}
\end{eqnarray}
where $\beta\leq 1$.
The conical singularity at $r=0$,
which corresponds to a codimension-2 tense brane,
is resolved by introducing
a cylindrical 4-brane at $r=r_0$ and filling in the interior with a regular cap.
As in~\cite{Kaloper}, we assume a disk-like cap whose metric is given by
\begin{eqnarray}
g_{AB}dx^Adx^B = \eta_{\mu\nu}dx^{\mu}dx^{\nu}+d\rho^2+\rho^2d\varphi^2
\nonumber\\
(\rho<\rho_0=\beta r_0),
\label{capmetric}
\end{eqnarray}
where $\rho_0$ is the position of the brane seen from the cap side.
In other words, $\rho_0$ is the size of the cap.
The continuity of the induced metric on the brane implies that $\rho_0=\beta r_0$.
Throughout the paper we assume that both the bulk and the cap are described by
6D Einstein gravity without a cosmological constant nor any other fields.
Both~(\ref{bulkmetric}) and~(\ref{capmetric}) solve the 6D vacuum Einstein equations trivially.
The two spacetimes are glued together along the ring-like brane, on which now
it is possible to put an arbitrary energy-momentum tensor.
The 4-brane action we consider is
\begin{eqnarray}
S_{\text{brane}} =
\int d^5x\sqrt{-q}\left(-\lambda-\frac{1}{2}q^{ab}\partial_a\Sigma\partial_b\Sigma+{\cal L}_{{\rm m}}\right),
\end{eqnarray}
where
$q_{ab}$ is the induced metric on the brane,
$\lambda$ is the tension of the brane, $\Sigma$ is the brane-localized scalar field,
and ${\cal L}_{{\rm m}}$ is the Lagrangian of usual matter. 
As we are describing the background configuration of the model,
we do not consider the contribution from ${\cal L}_{{\rm m}}$ in this section.
The background equation of motion, $\partial_{\varphi}\partial^{\varphi}\Sigma=0$, implies that
$\Sigma = Q \varphi$, where $Q$ is a constant.
The solution for $\Sigma$
breaks the translational invariance in the fifth direction on the brane, but
the energy-momentum tensor does not. This field is introduced
so as to cancel the pressure in the $\varphi$ direction coming from $\lambda$.

Note that
although we discuss only the very simple bulk geometry presented above,
such a regularization scheme works as well
in more general setups in which the bulk contains e.g.
fluxes and dilaton fields and is curved~\cite{Peloso, Papa, Reg-TK, UVcaps}.

Suppose now that the Planck scale in the cap region, $M_\rml$,
differs from that in the bulk, $M_\rmr$.
The junction conditions on the brane are then given by~\cite{Is}
\begin{eqnarray}
M^4_\rml \left.k_{ab}\right|_{ \rho_0}
-M^4_\rmr \left.k_{ab}\right|_{ r_0}={\cal T}_{ab},
\end{eqnarray}
where $k_{ab}=K_{ab}-q_{ab}K_c^{\;c}$, $K_{ab}$ is the extrinsic curvature of the brane,
\begin{eqnarray}
{\cal T}_{ab}=
-\lambda q_{ab}+\partial_{a}\Sigma\partial_{b}\Sigma
-\frac{1}{2}q_{ab}\partial_c\Sigma\partial^{c}\Sigma
+T_{ab},
\end{eqnarray}
and $T_{ab}$ is the energy-momentum tensor derived from ${\cal L}_{{\rm m}}$.
The background junction conditions read
\begin{eqnarray}
(\varphi\varphi): &&\;\; \lambda=\frac{1}{2}q^{\varphi\varphi}Q^2,\label{bj1}
\\
(\mu\nu):&&\;\;
\frac{M_\rml^4}{\rho_0}-\frac{M_\rmr^4}{r_0}=\lambda+\frac{1}{2}q^{\varphi\varphi}Q^2,\label{bj2}
\end{eqnarray}
with $q^{\varphi\varphi}=\rho_0^{-2}=(\beta r_0)^{-2}$.
Therefore, the brane tension satisfies
the relation $2\lambda=M_\rml^4/\rho_0-M_\rmr^4/r_0$.

We thus have defined the background configuration of the model (Fig.~1).
In contrast to the model of~\cite{Kaloper},
we do not introduce an induced gravity term in the 4-brane action explicitly.
Rather, we realize the regularization of brane-induced gravity proposed by~\cite{Reg-DGP},
using the capped bulk geometry.
The key assumption here is the very large Planck scale in the cap: $M_\rml\gg M_\rmr$.
The idea is also closely related to the work of~\cite{Padilla, KoyamaKoyama},
in which the asymmetry between
two sides of the brane gives rise to the effect analogous to brane-induced gravity.
The authors of~\cite{Padilla, KoyamaKoyama} considered the Randall-Sundrum-type
braneworlds, but the mechanism will operate also in the present setup.
In the rest of the paper,
we will show by a detailed perturbation analysis that 4D gravity indeed emerges
in a certain region of distance scales.

\begin{figure}[tb]
  \begin{center}
    \includegraphics[keepaspectratio=true,height=50mm]{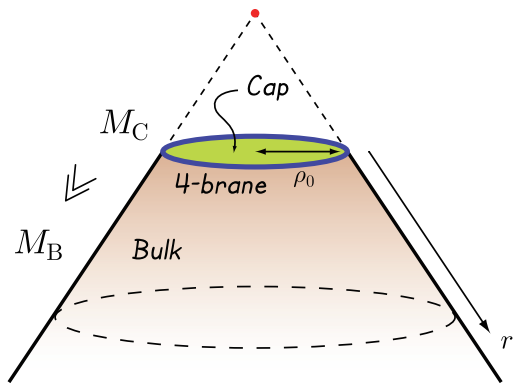}
  \end{center}
  \caption{The configuration of the model.}%
  \label{fig:cone.eps}
\end{figure}

\section{Linear analysis}\label{sec:lin}

\subsection{General analysis}\label{genan}

We study linearized gravity sourced by arbitrary matter on the brane.
The perturbed metric is given by
\begin{eqnarray}
&&(g_{AB}+\delta g_{AB})dx^Adx^B
=\left( \eta_{\mu\nu}+\gamma_{\mu\nu}\right)dx^{\mu}dx^{\nu}
\nonumber\\
&&\quad
+2B_{,\mu}dx^{\mu}d\rho+ (1+2\Gamma)d\rho^2+(1-2\Phi)\rho^2d\varphi^2,
\label{percap}
\end{eqnarray}
in the cap and
\begin{eqnarray}
&&(g_{AB}+\delta g_{AB})dx^Adx^B=
\left(\eta_{\mu\nu}+\gamma_{\mu\nu}\right)dx^{\mu}dx^{\nu}
\nonumber\\
&&\;\;
+2B_{,\mu}dx^{\mu}dr+ (1+2\Gamma)dr^2+(1-2\Phi)\beta^2r^2d\varphi^2,
\label{perbulk}
\end{eqnarray}
in the bulk.
The perturbations are split into scalar, vector, and tensor modes under the Lorentz group in
the 4D coordinates, and so we write
\begin{eqnarray}
\gamma_{\mu\nu}:=2\Psi\eta_{\mu\nu}+2E_{,\mu\nu}+h_{\mu\nu},
\end{eqnarray}
where $h_{\mu\nu}$ is the transverse and traceless tensor perturbation:
$h_\mu^{\;\mu}=\partial_\nu h_{\mu}^{\;\nu}=0 $.
Vector perturbations are not included above and will not be considered throughout the main text,
because, as is explained in Appendix~\ref{app:vec},
they do not contribute to gravity on the brane.
In this paper we only consider axisymmetric perturbations.

In Eqs.~(\ref{percap}) and~(\ref{perbulk}) the perturbed components $\delta g_{\rho\varphi}$,
$\delta g_{r\varphi}$, and $\delta g_{\mu\varphi}$ are eliminated
with the aid of the gauge transformation $\varphi\to\varphi+\delta\varphi$.
We can further use the gauge transformation $\rho\to \rho+\delta\rho\;(r\to r+\delta r)$
and $x^{\mu}\to x^{\mu}+\delta x^{\mu}$.
The linearized Einstein equations will be solved easily by invoking the specific gauge in which $B=E=0$
(the longitudinal gauge).
Our master equations in the cap are
\begin{eqnarray}
h_{\mu\nu}''+\frac{1}{\rho}h_{\mu\nu}'+\Box h_{\mu\nu}=0
\label{master-tensor}
\end{eqnarray}
for the tensor perturbations and
\begin{eqnarray}
\Psi''+\frac{1}{\rho}\Psi'+\Box\Psi=0
\label{master-scalar}
\end{eqnarray}
for the scalar perturbations, where the primes denote derivatives with respect to $\rho$
and $\Box:=\eta^{\mu\nu}\partial_{\mu}\partial_{\nu}$.
The other scalar quantities are obtained from
\begin{eqnarray}
\left(\rho^2\Phi\right)'&=&3\rho^2\Psi'+2\rho\Psi,\label{rel-scalar1}
\\
\Gamma&=&\Phi-2\Psi.\label{rel-scalar2}
\end{eqnarray}
The derivation of these equations is deferred to Appendix~\ref{app:per}.
Eqs.~(\ref{master-tensor})--(\ref{rel-scalar2}) are for the cap,
but we have the equations for the bulk by replacing $\rho$ with $r$ in the above.

The general solutions to the bulk and cap perturbation equations are found to be
\begin{eqnarray}
h_{\mu\nu}&=&
\begin{cases}
\displaystyle{
\int\hat h_{\mu\nu\rml}(p)I_0(p\rho)e^{ip\cdot x}d^4p
}\;\;\;\text{(cap)}\\
\displaystyle{
\int\hat h_{\mu\nu\rmr}(p)K_0(pr)e^{ip\cdot x}d^4p
}\;\;\text{(bulk)}
\end{cases},\label{gen_h}
\\
\Psi&=&
\begin{cases}
\displaystyle{
\int\hat\psi_\rml(p)I_0(p\rho)e^{ip\cdot x}d^4p
}\quad\;\text{(cap)}\\
\displaystyle{
\int\hat\psi_\rmr(p)K_0(pr)e^{ip\cdot x}d^4p
}\quad\text{(bulk)}
\end{cases},\label{gen_psi}
\\
\Phi&=&
\begin{cases}
\displaystyle{
\int\hat\psi_\rml(p) {\cal I}(p\rho) e^{ip\cdot x}d^4p
}\quad\text{(cap)}\\
\displaystyle{
\int\hat\psi_\rmr(p) {\cal K}(pr) e^{ip\cdot x}d^4p
}\quad\text{(bulk)}
\end{cases},\label{gen_phi}
\end{eqnarray}
where
\begin{eqnarray}
{\cal I}(x):=3I_2(x)+\frac{2I_1(x)}{x},
\\
{\cal K}(x):=3K_2(x)-\frac{2K_1(x)}{x},
\end{eqnarray}
and $K_n$ and $I_n$ are the modified Bessel functions.
Note that $I_0(p\rho)$ and ${\cal I}(p\rho)$
are regular at the center of the disk, $\rho=0$,
and $K_0(pr)$ and ${\cal K}(pr)$ remain finite as $r\to\infty$.

The above choice of the gauge forces the brane to move from its original position:
\begin{eqnarray}
\rho_0\to\rho_0+\zeta_\rml(x),
\quad
r_0\to r_0+\zeta_\rmr(x).
\end{eqnarray}
In order to impose the boundary conditions at the brane, it is thus convenient to go to
the Gaussian normal gauge, in which the brane does not flutter.
The tensor perturbation $h_{\mu\nu}$ is invariant under
an infinitesimal coordinate transformation.
According to
the gauge transformation of the scalar perturbations summarized in Appendix~\ref{app:sc},
the Gaussian normal gauge quantities evaluated on the brane
are related to the perturbations in the longitudinal gauge as
\begin{eqnarray}
&&\overline{\Psi}=\Psi|_{\rho_0},\quad
\overline\Phi=\Phi|_{\rho_0}-\frac{\zeta_\rml}{\rho_0},\quad\overline{E}=0,
\quad\overline{\Psi}'=\Psi'|_{\rho_0},
\nonumber\\&&
\overline\Phi'=\Phi'|_{\rho_0}+\frac{1}{\rho_0}\Gamma|_{\rho_0}+\frac{1}{\rho_0^2}\zeta_\rml,
\quad \overline{E}'=-\zeta_\rml.
\label{gt}
\end{eqnarray}
Again, replacing $\rho$ and $\zeta_\rml$ with $r$ and $\zeta_{\rmr}$, respectively,
we obtain the equations for the bulk side.

Since the induced metric must be continuous across the brane, we impose
\begin{eqnarray}
\Psi|_{\rho_0}=\Psi|_{r_0}\label{con_psi}
\end{eqnarray}
and
\begin{eqnarray}
-\Phi|_{\rho_0}+\frac{\zeta_{\rml}}{\rho_0} = -\Phi|_{r_0}+\frac{\zeta_\rmr}{r_0}=: \phi(x).
\label{con_phi}
\end{eqnarray}

The perturbed extrinsic curvature is given by
$\delta K_{\mu\nu}=(1/2)\overline{\gamma}_{\mu\nu}'$
and $\delta K_{\varphi}^{\;\varphi}=-\overline{\Phi}'$.
Using the transformation rule~(\ref{gt}),
the junction conditions are written in terms of the longitudinal gauge quantities as
\begin{eqnarray}
&&M_\rml^4\left(\frac{1}{2}h_{\mu\nu}'|_{\rho_0}+
{\cal D}_{\mu\nu}\zeta_\rml\right)
\nonumber\\&&\;
-M_\rmr^4\left(\frac{1}{2}h_{\mu\nu}'|_{r_0}+
{\cal D}_{\mu\nu}\zeta_\rmr\right)
=T_{\mu\nu},
\label{junc_munu}
\end{eqnarray}
and
\begin{eqnarray}
&&M_\rml^4 \left(\Box\zeta_\rml-4 \Psi'|_{\rho_0}+\frac{\phi}{\rho_0}\right)
\nonumber\\&&\;
-M_\rmr^4 \left(\Box\zeta_\rmr-4 \Psi'|_{r_0}+\frac{\phi}{r_0}\right)
=T_{\varphi}^{\;\varphi},
\label{junc_phiphi}
\end{eqnarray}
where ${\cal D}_{\mu\nu}:=\eta_{\mu\nu}\Box-\partial_{\mu}\partial_{\nu}$.
We used the background junction conditions and the perturbed Einstein equations
$\delta G_{\mu\rho}=\delta G_{\mu r}=0$ to make the expressions as compact as possible.
(The derivation and details are presented in Appendix~\ref{app:der}.)
Note that the perturbation of the brane-localized scalar field, $\delta\Sigma(x)$,
does not appear in the junction conditions as long as axisymmetric perturbations are considered.
The 4D trace of the $(\mu\nu)$ component of the junction conditions
is useful:
\begin{eqnarray}
M_\rml^4\Box\zeta_\rml
-M_\rmr^4\Box\zeta_\rmr=\frac{1}{3}T_{\mu}^{\;\mu}.\label{tr_junc_munu}
\end{eqnarray}

We now define
\begin{eqnarray}
M_4^2=\pi\rho^2_0M_\rml^4,\quad r_c=\frac{\rho_0M_\rml^4}{2M_\rmr^4},
\end{eqnarray}
and the energy-momentum tensor integrated along the $\varphi$ direction: $\overline{T}_{ab}=2\pi\rho_0 T_{ab}$.
Then, the above junction conditions are rewritten as
\begin{eqnarray*}
&&M_\rml^4(\,\text{cap}\,)-M_\rmr^4(\,\text{bulk}\,)=T_{**}
\nonumber\\
&&\qquad\;\; \Leftrightarrow  \;\;
\frac{2}{\rho_0}(\,\text{cap}\,)- \frac{1}{r_c}(\,\text{bulk}\,)=\frac{\overline{T}_{**}}{M_4^2}.
\end{eqnarray*}

Recall now that the brane position seen from the bulk side, $r_0$,
is related to the size of the disk-like cap, $\rho_0$, as $r_0=\rho_0/\beta$,
where $\beta$ controls the opening of the cone.
We assume that $\rho_0\ll r_c\lesssim r_0$. This requires
$\beta\ll 1$ (i.e., a very narrow cone) and $M_\rml\gg M_\rmr$.
By $r_c\lesssim r_0$
we mean to allow for both $r_0<r_c$ and $r_c\ll r_0$, but we preclude $r_0\ll r_c$.
Let us focus on scales which lie between $\rho_0$ and $r_0$.
The approximation
$I_0(p\rho_0)\simeq1+(p\rho_0)^2/4$ yields
\begin{eqnarray}
\left.  h_{\mu\nu}'\right|_{\rho_0} &\simeq&
\frac{\rho_0}{2}\int\hat h_{\mu\nu\rml}(p)p^2e^{ip\cdot x}d^4p
\nonumber\\
&=&-\frac{\rho_0}{2}\left.\Box h_{\mu\nu}\right|_{\rho_0}.\label{h'=boxh}
\end{eqnarray}
Similarly, we have
\begin{eqnarray}
\left.  \Psi'\right|_{\rho_0} \simeq -\frac{\rho_0}{2}\Box\Psi|_{\rho_0}\label{p'=boxp}
\end{eqnarray}
and
\begin{eqnarray}
\Psi|_{\rho_0}\simeq\Phi|_{\rho_0}, \label{p=1p}
\end{eqnarray}
where we used ${\cal I}(p\rho_0)\simeq 1$ in deriving the second relation.
Noting that 4D Ricci tensor of the metric $\eta_{\mu\nu}+\overline{\gamma}_{\mu\nu}$
is given by
\begin{eqnarray}
{\cal R}_{\mu\nu}=-\frac{1}{2}\Box h_{\mu\nu}|_{\rho_0}
-\eta_{\mu\nu}\Box\Psi|_{\rho_0}-2\partial_{\mu}\partial_{\nu}\Psi|_{\rho_0},
\end{eqnarray}
we find
\begin{eqnarray}
\frac{2}{\rho_0}\left(\frac{1}{2}h_{\mu\nu}'|_{\rho_0}+
{\cal D}_{\mu\nu} \zeta_\rml \right)
\simeq   {\cal R}_{\mu\nu}-\frac{1}{2}\eta_{\mu\nu}{\cal R}
+2{\cal D}_{\mu\nu} \phi.\label{cap_R=}
\end{eqnarray}
As for the bulk side, we make the approximation $pr_0\gg 1$ and so
${\cal K}(pr_0)\approx 3K_0(pr_0)$, leading to
\begin{eqnarray}
\Phi|_{r_0}\simeq 3\Psi|_{r_0}.\label{p=3pbulk}
\end{eqnarray}
Using this and the perturbed Einstein equation $\delta G_{rr}=0$,
we find
\begin{eqnarray}
\Psi'|_{r_0}\simeq 0.\label{p'bulk}
\end{eqnarray}

Now Eqs.~(\ref{con_psi}),~(\ref{con_phi}),~(\ref{p=1p}), and~(\ref{p=3pbulk}) yield
\begin{eqnarray}
\zeta_\rml\simeq\rho_0\left(\phi+\Psi|_{\rho_0}\right),
\quad
\zeta_\rmr\simeq r_0\left(\phi+3\Psi|_{\rho_0}\right).
\label{zeta=p+p}
\end{eqnarray}
Substituting Eqs. (\ref{p'=boxp}),~(\ref{p'bulk}), and~(\ref{zeta=p+p})
into the junction conditions~(\ref{junc_phiphi}) and~(\ref{tr_junc_munu})
and solving for $\phi$ and $\Psi|_{\rho_0}$,
we obtain
\begin{eqnarray}
(3-2\alpha)\phi&\simeq& -\frac{\rho_0^2}{2M_4^2}\left(\overline{T}_{\mu}^{\;\mu}-
3\overline{T}_{\varphi}^{\;\varphi}\right)
\nonumber\\&&\quad
+\frac{\alpha\rho_0^2}{M_4^2}\left(\overline{T}_{\mu}^{\;\mu}-
\overline{T}_{\varphi}^{\;\varphi}\right),\label{sol-phi}
\\
(3-2\alpha)\Psi&\simeq&\frac{\rho_0^2}{6M_4^2}\left(\overline{T}_{\mu}^{\;\mu}-
3\overline{T}_{\varphi}^{\;\varphi}\right)
\nonumber\\&&\quad
-\frac{\alpha}{3M_4^2}\Box^{-1}
\overline{T}_{\mu}^{\;\mu}, \label{sol-psi}
\end{eqnarray}
with $\alpha:=r_c/r_0$.
Here, we
neglected terms suppressed by factors $\rho_0^2\Box$, $\rho_0^2/r_0^2$,
and $\rho_0^2/r_c^2$ relative to the others. However,
we keep the first term in the right hand side of Eq.~(\ref{sol-psi})
because $\alpha$ may also be small.
From Eq.~(\ref{sol-phi}) one finds that $\phi$ is algebraically
determined by the energy-momentum tensor on the brane and hence is a nonpropagating mode.
In what follows we do not consider the case in which $\alpha$ is equal to or very close to $3/2$.
Plugging Eqs.~(\ref{sol-phi}) and~(\ref{sol-psi}) into the expression~(\ref{zeta=p+p}),
we can write
the brane bending scalars in terms of the energy-momentum tensor as
\begin{eqnarray}
(3-2\alpha)\frac{\zeta_{\rml}}{\rho_0}&\simeq& -\frac{\rho_0^2}{3M_4^2}\left(\overline{T}_{\mu}^{\;\mu}-
3\overline{T}_{\varphi}^{\;\varphi}\right)
\nonumber\\&&\quad
-\frac{\alpha}{3M_4^2}\Box^{-1}\overline{T}_{\mu}^{\;\mu},
\\
(3-2\alpha)\frac{\zeta_{\rmr}}{r_c}&\simeq&-\frac{1}{M_4^2}\Box^{-1}\overline{T}_{\mu}^{\;\mu}.
\label{sol-zb}
\end{eqnarray}

Using Eq.~(\ref{cap_R=}),
the junction conditions~(\ref{junc_munu}) are now written as
\begin{eqnarray}
{\cal R}_{\mu\nu}-\frac{1}{2}\eta_{\mu\nu}{\cal R}
\simeq\frac{\overline{T}_{\mu\nu}}{M_4^2}
-{\cal D}_{\mu\nu}\chi
+\frac{h_{\mu\nu}'|_{r_0}}{2r_c},\label{effeq}
\end{eqnarray}
where $\chi:=-\zeta_\rmr/r_c$.
Here, the term
${\cal D}_{\mu\nu}\phi$
was neglected because Eq.~(\ref{sol-phi}) implies that it is smaller
by a factor of $\rho_0^2\Box$ than the other terms.
Since $\phi=\overline{\delta q}_{\varphi\varphi}/2$,
this suppressed mode corresponds to the fluctuation of the size of the ring-like brane.
The same suppression is found also in~\cite{Peloso, Reg-TK, Takamizu}.
As $K_0'(pr)|_{r_0}\approx-pK_1(pr_0)\approx-pK_0(pr_0)$ for $pr_0\gg 1$,\footnote{Noting
that $h_{\mu\nu}'|_{r_0}\simeq [\overline{\gamma}_{\mu\nu}'-\eta_{\mu\nu}\overline{\gamma}']|_{r_0}$, the
last term in Eq.~(\ref{effeq}) implies Pauli-Fierz type massive gravity~\cite{PFterm} with
$m^2_{{\rm PF}}=\sqrt{-\Box}/r_c$, as in the 5D DGP model.
}
we have the estimate
\begin{eqnarray}
\frac{1}{r_c}h_{\mu\nu}'|_{r_0}\sim\frac{1}{r_c\mathtt{r}}h_{\mu\nu}|_{r_0},\label{dgpterm}
\end{eqnarray}
where $\mathtt{r}$ is the length scale under consideration,
so that the last term in the right hand side of Eq.~(\ref{effeq}) is negligible when $\mathtt{r}\ll r_c$.
The situation here is very similar to that of the DGP braneworld~\cite{DGP}.
From Eqs.~(\ref{effeq}) and~(\ref{sol-zb}), i.e.,
\begin{eqnarray}
\Box\chi\simeq \frac{1}{3-2\alpha}\frac{\overline{T}_{\mu}^{\;\mu}}{M_4^2},
\label{chitrace}
\end{eqnarray}
we conclude that linearized
gravity on scales smaller than $r_c$ and $r_0$
is effectively described by the 4D Brans-Dicke theory
with $\omega_{{\rm BD}}=-r_c/r_0$~\cite{BD}, and the Brans-Dicke scalar $(\chi)$ is identified
as the brane bending.
In order for this scalar not to be a ghost,
it is required that $3-2\alpha>0$~\cite{FujiiMaeda}.
Therefore, we preclude the case with $r_c\geq 3r_0/2\, (\gg\mathtt{r}\gg \rho_0)$.

As the volume of the bulk is infinite in the present model,
the zero mode is not normalizable and hence is removed from the spectrum.
Therefore, from the brane observer point of view,
Kaluza-Klein modes contribute to gravity on the brane.
Nevertheless, 4D gravity is recovered
because nearly massless modes are confined around the brane in the cap side.
These modes
play the role of 5D induced gravity term ``$R^{(5)}$'' with one dimension compactified to a small circle.
The fluctuation of the size of the Kaluza-Klein circle is suppressed, and hence
we can expect to recover a 4D tensor structure.

We may expect that gravity shows the higher dimensional nature
at longer distances, $\mathtt{r}\gtrsim r_c, r_0$.
From the higher dimensional perspective,
gravitons leak away into the extra dimensions at distances much larger than $r_c$.
However, for $r_c\ll\mathtt{r}\ll r_0$ one of the extra dimensions is
effectively compactified because the extra 2D space is just like a needle,
so that gravity will look 5D~\cite{Kaloper1}.
Working in the momentum space, one sees that gravity is indeed 5D because
the differentiation with respect to the bulk coordinate $r$
yields $p$ rather than $p^2$
and the last term in Eq.~(\ref{effeq}) is much bigger than ${\cal O}({\cal R}_{\mu\nu})$
in this regime.
On much larger scales, $\mathtt{r}\gg r_0$, a graviton will see the opening of the cone,
leading to the 6D behavior.
This picture is indeed true, as is shown by an explicit example in the next subsection.

Moreover, this is not the end of the story because,
even in the weak gravity regime, nonlinearities will set in below
a certain distance scale.
Although the Brans-Dicke behavior contradicts tests of gravity,
nonlinear effects help to remove the scalar degree of
freedom,
leaving 4D Einstein gravity on
the brane~\cite{Deffayet, Lue:2001gc, Gabadadze:2004iy, Gruzinov:2001hp, Tanaka, KScp}.
This is what will be discussed in Sec.~\ref{sec:nonlinear}.

\subsection{Case study: static ``point'' sources}\label{sec:pointsource}

In this subsection we compute the gravitational field produced by
a static point source.
Concerning gravity at short distances,
this is just a check of our general argument in the previous subsection.
This example also allows us to see explicitly how
gravity is modified at long distances.
Here we restrict ourselves to linear perturbations.


As in~\cite{Kaloper}, we model a static point source in our 4D world
as a loop of string along the ring-like brane.
The energy momentum tensor for a static string loop is given by
\begin{eqnarray}
\overline{T}_t^{\;t}=\overline{T}_\varphi^{\;\varphi}=
-{\cal M}\;\delta^{(3)}(\vec{x}),\quad
\overline{T}_{ij}=0,
\end{eqnarray}
with $i, j$ being 3-space indices.
Since the source and the gravitational field are static, we use the expression
\begin{eqnarray}
h_{\mu\nu}&=&
\begin{cases}
\displaystyle{
\int\hat h_{\mu\nu}(k)K_0(kr_0)I_0(k\rho)e^{i\vec{k}\cdot \vec{x}}d^3k
}\;\;\;\text{(cap)}\\
\displaystyle{
\int\hat h_{\mu\nu}(k)I_0(k\rho_0)K_0(kr)e^{i\vec{k}\cdot \vec{x}}d^3k
}\;\;\text{(bulk)}
\end{cases},\label{genk_h}
\\
\Psi&=&
\begin{cases}
\displaystyle{
\int\hat\psi(k)K_0(kr_0)I_0(k\rho)e^{i\vec{k}\cdot \vec{x}}d^3k
}\quad\;\text{(cap)}\\
\displaystyle{
\int\hat\psi(k)I_0(k\rho_0)K_0(kr)e^{i\vec{k}\cdot \vec{x}}d^3k
}\quad\text{(bulk)}
\end{cases},\label{genk_psi}
\\
\Phi&=&
\begin{cases}
\displaystyle{
\int\hat\psi (k) K_0(kr_0){\cal I}(k\rho) e^{i\vec{k}\cdot \vec{x}}d^3k
}\quad\text{(cap)}\\
\displaystyle{
\int\hat\psi(k)I_0(k\rho_0) {\cal K}(kr) e^{i\vec{k}\cdot \vec{x}}d^3k
}\quad\text{(bulk)}
\end{cases},\label{genk_phi}
\end{eqnarray}
where the continuity of $h_{\mu\nu}$ and $\Psi$ has been implemented.
In the following calculation, it is convenient to use the 4D Green function:
\begin{eqnarray}
{\cal G}_4(x)=\frac{1}{4\pi|\vec{x}|},\quad
\nabla^2{\cal G}_4 = -\delta^{(3)}(\vec{x}),
\end{eqnarray}
where $\nabla^2:=\delta^{ij}\partial_i\partial_j$.

The trace of the $(\mu\nu)$
junction conditions~(\ref{tr_junc_munu}) reads
\begin{eqnarray}
2\frac{\zeta_\rml}{\rho_0}-\frac{\zeta_\rmr}{r_c}=\frac{{\cal M}}{3M_4^2}{\cal G}_4.
\label{2z-z}
\end{eqnarray}
Plugging this into~(\ref{junc_munu}), we obtain
\begin{eqnarray}
K_0(kr_0)\hat h_{00}(k)&\simeq& \frac{4{\cal M}}{3M_4^2}\hat{\cal G}(k),
\\
K_0(kr_0)\hat h_{ij}(k)&\simeq& \frac{2{\cal M}}{3M_4^2}
\left(\delta_{ij}-\frac{k_ik_j}{k^2}\right)\hat{\cal G}(k),
\end{eqnarray}
where
\begin{eqnarray}
\hat{\cal G}(k):=\frac{1}{(2\pi)^3}\left[k^2+\frac{k}{r_c}\frac{K_1(kr_0)}{K_0(kr_0)}\right]^{-1},
\label{kernel-G}
\end{eqnarray}
and we made use of the small argument expansion $I_0(k\rho_0)\simeq{\cal I}(k\rho_0)\simeq 1$
and $I_0'(k\rho)|_{\rho_0}\simeq k^2\rho_0/2$.
From the continuity of $\overline{\Phi}$ [Eq.~(\ref{con_phi})] we find
\begin{widetext}
\begin{eqnarray}
\left(1-2\alpha\right)\phi
=-\alpha\frac{{\cal M}}{3M_4^2}  {\cal G}_4
+\int\hat\psi(k)\left[2\alpha K_0(kr_0)-{\cal K}(kr_0)\right]e^{i\vec{k}\cdot\vec{x}}d^3k,
\end{eqnarray}
The $(\varphi\varphi)$ component of the junction equations now reads
\begin{eqnarray}
\int\hat\psi(k)\left\{-\left(1-2\alpha\right)\left[k^2K_0(kr_0)+\frac{kK_1(kr_0)}{r_c}\right]
+\alpha \frac{K_0(kr_0)}{\rho_0^2}-\frac{{\cal K}(kr_0)}{2\rho_0^2}
\right\}e^{i\vec{k}\cdot\vec{x}}d^3k
\simeq\frac{{\cal M}}{6M_4^2}\left[
\frac{\alpha}{\rho_0^2}
+\left(1-2\alpha\right) \nabla^2\right]{\cal G}_4,\label{fourierJ}
\end{eqnarray}
\end{widetext}
where we neglected $\phi/(r_0r_c)\;(\ll\phi/\rho_0^2)$.
Looking at the small argument expansion
$K_0(x)\simeq -\gamma-\ln(x/2)$, $K_1\simeq 1/x$, and ${\cal K}(x)\simeq 4/x^2$ with
$\gamma=0.5772...\,$,
it turns out that the last two terms in the left hand side of Eq.~(\ref{fourierJ})
are much larger than the first two when $kr_0\ll 1$.
When $kr_0\gg 1$, the last two terms are again dominant.
Therefore, we have
\begin{eqnarray}
\left[2\alpha K_0(kr_0)-{\cal K}(kr_0)\right]\hat\psi\simeq
\frac{1}{(2\pi)^3}\frac{{\cal M}}{3M_4^2}\left(\frac{\alpha}{k^2}-\rho_0^2\right).
\label{kernel-psi}
\end{eqnarray}
Using this, it is easy to find $\phi\simeq 0$ and
\begin{eqnarray}
\frac{\zeta_\rml}{\rho_0}\simeq\Psi|_{\rho_0},\quad
\frac{\zeta_\rmr}{r_c}\simeq 2\Psi|_{\rho_0}-\frac{{\cal M}}{3M_4^2}{\cal G}_4,
\label{ps_z}
\end{eqnarray}
where note that the last term in the right hand side of Eq.~(\ref{kernel-psi})
gives a delta function in the real space, which vanishes away from the source.


\subsubsection{$|\vec{x}|\ll r_0$}

This is nothing but the case studied in Sec.~\ref{genan}.
As we may approximate $K_0(kr_0)\approx K_1(kr_0)$ in this case,
we have
\begin{eqnarray}
\hat{\cal G}(k)\simeq \frac{1}{(2\pi)^3}\frac{1}{k^2+k/r_c}.
\end{eqnarray}
In the real space, the behavior of $h_{\mu\nu}$ is governed by
\begin{eqnarray}
\int\hat{\cal G}(k)e^{i\vec{k}\cdot\vec{x}}d^3k&\simeq&
\begin{cases}
{\cal G}_4(x)\quad\quad\quad\quad (|\vec{x}|\ll r_c)
\\
r_c/(2\pi^2|\vec{x}|^2) \quad\; (|\vec{x}|\gg r_c)
\end{cases},
\end{eqnarray}
i.e., this reduces to ${\cal G}_4$ at short distances and shows 5D behavior for $r_c\ll |\vec{x}|\ll r_0$.
As for the scalar-type perturbations,
we make the approximation
$\left[2\alpha K_0(kr_0)-{\cal K}(kr_0)\right]\hat\psi\approx- (3-2\alpha)K_0(kr_0)\hat\psi$
to obtain
\begin{eqnarray}
\Psi|_{\rho_0}\simeq
-\frac{\alpha}{3-2\alpha}\frac{{\cal M}}{3M_4^2}{\cal G}_4(x).\label{r<r0_psi}
\end{eqnarray}
Thus, gravity is 4D at short distances, $|\vec{x}|\ll r_c$, in agreement with the argument in Sec.~\ref{genan}.
It is easy to confirm that the parameterized post-Newtonian (PPN) parameter is given by that of
the Brans-Dicke theory with $\omega_{{\rm BD}}=-\alpha=-r_c/r_0$.
Despite the 4D behavior of $\Psi|_{\rho_0}$,
this is always suppressed relative to $h_{\mu\nu}$ if $r_c\ll r_0$.
Therefore, for $r_c\ll |\vec{x}|\ll r_0$
the dominant contribution comes from $h_{\mu\nu}$
and gravity on the brane looks 5D.
If $r_c\sim r_0$, there is no 5D regime from the beginning.

\subsubsection{$|\vec{x}|\gg r_0 \;(\gtrsim r_c)$}\label{psA2}

In this regime we have ${\hat {\cal G}}(k)\simeq(2\pi)^{-3}r_0r_cK_0(kr_0)$,
and hence\footnote{
We use the following formula:
$$
\int_0^{\infty}y K_0(y)\sin(\alpha y)dy=\frac{\pi}{2}\frac{\alpha^{-2}}{(1+\alpha^{-2})^{3/2}}
\quad (\alpha>0).
$$
}
\begin{eqnarray}
h_{00}\simeq\frac{4{\cal M}}{3M_4^2} {\cal G}_6,
\;\;
h_{ij}\simeq \frac{2{\cal M}}{3M_4^2}
\left(\delta_{ij}-\partial_i\partial_j\nabla^{-2}\right){\cal G}_6,
\end{eqnarray}
where ${\cal G}_6(x):=r_0r_c/(4\pi|\vec{x}|^3)$.
Similarly, we find
$\hat\psi(k)\simeq-(2\pi)^{-3}r_0r_c ({\cal M} /12M_4^2)$,
leading to $\Psi|_{\rho_0}\simeq-({\cal M}/12M_4^2){\cal G}_6$.
Thus, we conclude that on the largest distance scales
gravity on the brane looks 6D.

\section{(Semi)nonlinear regime}\label{sec:nonlinear}

Based on the linear analysis, we have shown that for $\mathtt{r}\ll r_0$
the brane bending in the bulk side is given by
\begin{eqnarray}
\zeta_\rmr\sim r_c \frac{R_g}{\mathtt{r}},\label{z-rc}
\end{eqnarray}
where $R_g$ is the gravitational radius of the source [see Eq.~(\ref{sol-zb})]. (We
dropped the factor $(3-2\alpha)$ because we are assuming that
this always gives an ${\cal O}(1)$ coefficient.)
Since $r_c$ is a ``large'' parameter, nonlinearity of $\zeta_\rmr$ may be important.
Indeed, when
\begin{eqnarray}
(\partial_{\mu}\zeta_\rmr)^2\sim \frac{R_g}{\mathtt{r}},\label{zz-rgr}
\end{eqnarray}
such nonlinear terms in the brane bending must be taken into account
even if gravity is weak, $R_g/\mathtt{r}\ll 1$.
Substituting Eq.~(\ref{z-rc}) into Eq.~(\ref{zz-rgr}) and using the estimate $\partial_{\mu}\sim\mathtt{r}^{-1}$,
we find that nonlinear terms become important when
\begin{eqnarray}
\mathtt{r} \lesssim R_*:= (r_c^2 R_g)^{1/3}. 
\end{eqnarray}
The linear analysis in the previous section is reliable only for $\mathtt{r}\gtrsim R_*$.
In this section, we present a preliminary analysis of
weak gravity on even smaller scales, $(\rho_0\ll)\;\mathtt{r}\ll R_*$, taking into account
quadratic terms in $\zeta_\rmr$~\cite{Padilla, Lue:2001gc, Tanaka}.
Note that since $\zeta_\rml$ is estimated to be
$\sim$ max$\left\{(\rho_0/\mathtt{r})^3R_g, (r_c/r_0)(\rho_0/\mathtt{r})R_g\right\}$
according to the linear analysis,
nonlinear terms in $\zeta_\rml$ will never become relevant.
We are considering the weak gravity regime,
so that we will keep linear terms in the metric perturbations.


Although we are now
working in the quadratic terms in $\zeta_\rmr$,
we may still use the 6D linearized equations in the bulk.
Therefore,
the relations~(\ref{p=3pbulk}) and~(\ref{p'bulk}), 
which were derived
only by using the general solution to the bulk Einstein equations,
hold as well as
the cap equations~(\ref{h'=boxh})--(\ref{cap_R=}).
However,
we must treat more carefully
the transformation from the longitudinal gauge to
the Gaussian normal gauge in the bulk side when we impose the boundary conditions.

We move from the longitudinal gauge to the Gaussian normal gauge by
making the infinitesimal coordinate transformation,
$x^{\mu}=\overline{x}^{\mu}-\delta x^{\mu}(\overline{x}, \overline{r})$,
$r=\overline{r}-\delta r(\overline{x}, \overline{r})$,
satisfying the conditions
\begin{eqnarray}
\delta x^{\mu}|_{r_0}=0 \quad\text{and}\quad \delta r|_{r_0}=-\zeta_\rmr(x).
\end{eqnarray}
Under the above transformation, the metric perturbation transforms as
\begin{eqnarray}
&&\overline{\delta g}_{AB}(\overline{x}, \overline{r})\simeq
\delta g_{AB}(\overline{x}-\delta x, \overline{r}-\delta r)
\nonumber\\&&\quad
-g_{AB,r}\delta r+\frac{1}{2}g_{AB,rr}(\delta r)^2
\nonumber\\&&\quad\quad
-\delta x^{C}_{\;,A}g_{CB}(\overline{x}-\delta x, \overline{r}-\delta r)-(A\leftrightarrow B)
\nonumber\\&&\quad\quad\quad
+\delta x^{C}_{\; ,A}\delta x^{D}_{\;,B}g_{CD}(\overline{x}-\delta x, \overline{r}-\delta r),
\label{2nd_trans}
\end{eqnarray}
where we neglected terms like ${\cal O}(\delta x^C_{\;,A})\times{\cal O}(\delta g_{BC})$.
Since $B=\overline{B}=\overline{\Gamma}=0$,
we require
\begin{eqnarray}
0&=&-\delta x_{\mu, r}-\delta r_{,\mu}+\delta r_{,\mu}\delta r_{,r}+\delta x^{\nu}_{\;,\mu}\delta x_{\nu,r},
\\
0&=&2\Gamma-2\delta r_{,r} +(\delta r_{,r})^2+\delta x_{\mu,r}\delta x^{\mu}_{\;,r}.
\end{eqnarray}
Evaluating these two equations at $\overline{r}=r_0$, we obtain
\begin{eqnarray}
\left. \delta x_{\mu,r}\right|_{r_0}&=&\partial_{\mu}\zeta_\rmr,
\label{d_rx2}
\\
\left.\delta r_{,r}\right|_{r_0}&=&\Gamma|_{r_0}+\frac{1}{2}\partial_{\mu}\zeta_\rmr\partial^{\mu}
\zeta_\rmr.
\label{d_rr2}
\end{eqnarray}
Using Eqs.~(\ref{2nd_trans}),~(\ref{d_rx2}), and~(\ref{d_rr2}), we can compute the
metric perturbations and their $r$ derivatives evaluated on the brane:
\begin{eqnarray}
&&\overline{\Phi}=\Phi|_{r_0}-\frac{\zeta_\rmr}{r_0}-\frac{\zeta_\rmr^2}{2r_0^2},
\quad
\overline{\gamma}_{\mu\nu}=\gamma_{\mu\nu}|_{r_0}+\partial_\mu\zeta_\rmr\partial_{\nu}\zeta_\rmr,
\nonumber\\
&&\overline{\Phi}'=\Phi'|_{r_0}+\frac{1}{r_0}\Gamma|_{r_0}+\frac{\zeta_\rmr}{r_0^2}+\frac{1}{2r_0}
\partial_{\mu}\zeta_\rmr\partial^{\mu}\zeta_\rmr+\frac{\zeta_\rmr^2}{r_0^3},
\nonumber\\
&&\overline{\gamma}_{\mu\nu}'=\gamma_{\mu\nu}'|_{r_0}-2\partial_{\mu}\partial_{\nu}\zeta_\rmr.
\label{trmet2}
\end{eqnarray}
The brane bending scalar associated with $\partial_{\mu}$ will become large at short distances,
but we may assume that $\zeta_\rmr/r_0$ and $\zeta_\rmr/r_c$
remain small (i.e., order of metric perturbations) even in the seminonlinear regime.
We therefore neglect $\zeta_\rmr^2/r_0^2$ in the first equation
and $\zeta_\rmr^2/r_0^3$ in the third equation.

Now the $(\mu\nu)$ component of the junction conditions reduces to
\begin{eqnarray}
&&{\cal R}_{\mu\nu}-\frac{1}{2}\eta_{\mu\nu}{\cal R}
-\frac{\overline{T}_{\mu\nu}}{M_4^2}
\nonumber\\&&\;=-2{\cal D}_{\mu\nu}\phi
+\frac{1}{r_c} {\cal D}_{\mu\nu}\zeta_\rmr 
+\frac{1}{2r_cr_0}
\partial_\lambda\zeta_\rmr
\partial^{\lambda}\zeta_\rmr\eta_{\mu\nu}, 
\label{j2_mn}
\end{eqnarray}
where the last term in the right hand side is in fact negligible
because $\zeta_\rmr/r_c\gg (\zeta_\rmr/r_0)(\zeta_\rmr/r_c)$.
Thus, even though we
have taken into account the possible nonlinear terms
in the brane bending,
the $(\mu\nu)$ junction conditions turn out to be
the same as the linear result.
Note that we have already neglected the term $h_{\mu\nu}'|_{r_0}/r_c$
which is much smaller than $\Box h_{\mu\nu}|_{\rho_0}$.
The 4D trace of the $(\mu\nu)$ junction conditions reduces to
$
2\Box\zeta_\rml/\rho_0-\Box\zeta_\rmr /r_c=
\overline{T}_{\mu}^{\;\mu}/(3M_4^2),
$
neglecting the nonlinear term for the reason stated above.
The $(\varphi\varphi)$ component of the junction conditions is given by
\begin{eqnarray}
\frac{2}{\rho_0}\left(\Box\zeta_\rml+2\rho_0\Box\Psi|_{\rho_0}+\frac{\phi}{\rho_0}\right)
-\frac{1}{r_c}\Box\zeta_\rmr
=\frac{\overline{T}_{\varphi}^{\;\varphi}}{M_4^2},
\label{junc_phiphi_nl}
\end{eqnarray}
where we used
Eqs.~(\ref{p'=boxp}) and~(\ref{p'bulk}), and
dropped the term $\phi/(r_0r_c)$ which is much smaller than $\phi/\rho_0^2$.
Again, Eq.~(\ref{junc_phiphi_nl}) is the same as the linear result.

The continuity of the induced metric reads
$\phi\simeq-\Psi|_{\rho_0}+\zeta_\rml/\rho_0\simeq-3\Psi|_{r_0}+\zeta_{\rmr}/r_0$.
We use the trace of the second equation in~(\ref{trmet2}),
$\Psi|_{\rho_0}=\Psi|_{r_0}+\partial_{\mu}\zeta_\rmr\partial^{\mu}\zeta_{\rmr}/8$,
to get
\begin{eqnarray}
\frac{\zeta_\rml}{\rho_0}
-\frac{1}{3}\frac{\zeta_\rmr}{r_0}-\frac{2}{3}\phi
\simeq\frac{1}{8}\partial_{\mu}\zeta_\rmr\partial^{\mu}\zeta_{\rmr}.\label{quadra}
\end{eqnarray}
The nonlinear effect appears here and this will play a very important role
in screening away the scalar degree of freedom $\chi$ found in the linear analysis.

Suppose that the nonlinear effect dominates in the above equations:
$(\partial_{\mu}\zeta_\rmr)^2\gg\zeta_\rmr/r_c, \zeta_\rmr/r_0$.
In this limit, one finds
\begin{eqnarray}
\frac{\zeta_\rml}{\rho_0}&\simeq&\frac{1}{6M_4^2}\Box^{-1}\overline{T}_{\mu}^{\;\mu}
\,\simeq\,\Psi|_{\rho_0},
\\
\phi&\simeq& \frac{\rho_0^2}{2M_4^2}\left(\overline{T}_{\varphi}^{\;\varphi}
-\overline{T}_{\mu}^{\;\mu}\right),\label{2ndphi}
\end{eqnarray}
and
\begin{eqnarray}
\partial_{\mu}\zeta_\rmr\partial^{\mu}\zeta_{\rmr}\simeq\frac{4}{3M_4^2}
\Box^{-1} \overline{T}_{\mu}^{\;\mu}.\label{2ndz2}
\end{eqnarray}
Because of the large contribution from the quadratic term in Eq.~(\ref{quadra}),
the result obtained here is different from the linear calculation~(\ref{sol-phi})--(\ref{sol-zb}).
From Eq.~(\ref{2ndz2}) we obtain the estimate $\zeta_\rmr\sim\sqrt{R_g\mathtt{r}}$.
For $\mathtt{r}\ll R_*$ we indeed have
$(\partial_{\mu}\zeta_\rmr)^2\sim$ (metric perturbations) $\gg\zeta_\rmr/r_c$,
and so the approximation is self-consistent.
Eq.~(\ref{2ndphi}) indicates that $\phi$ is suppressed
in much the same way as in the linear analysis.
Since ${\cal D}_{\mu\nu}\zeta_\rmr/r_c\ll{\cal O}({\cal R}_{\mu\nu})$,
all the terms in the right hand side of Eq.~(\ref{j2_mn}) are now found to be negligible.
Consequently, {\em 4D Einstein gravity is recovered on scales much smaller than $R_*$.}
In other words, the van Dam-Veltman-Zakharov discontinuity~\cite{vDVZ} disappears.


The analysis so far has neglected terms like
$ \gamma_{\mu\nu}\partial^{\nu}\zeta_\rmr$, $\gamma_{\mu\nu,r}\zeta_\rmr$,
$\gamma_{\mu\nu,rr}\zeta_\rmr^2, ...\,$.
Noting that $(\partial_r)^{n}\delta g_{AB}|_{r_0}\sim \delta g_{AB}|_{r_0}/\mathtt{r}^n$
and $\partial_{\mu}\zeta_\rmr\sim\zeta_\rmr/\mathtt{r}\sim\sqrt{R_g/\mathtt{r}}$,
one finds that such terms give higher order contributions.

\section{Summary and Conclusions}

Summarizing, we have studied weak gravity
in the 6D regularized braneworld having
the mechanism of long distance modification of gravity.
We resolved the codimension-2 singularity
by capping the apex of the cone-shaped bulk
and replacing the conical brane with a ring-like codimension-1 brane.
The brane contains a scalar field which cancels the pressure along the compact direction.
In regularizing the codimension-2 brane,
we have assumed the Planck scale in the cap to be much greater than the bulk Planck scale.
This assumption is along the line of the regularization of brane-induced gravity of~\cite{Reg-DGP}.
We have shown that the model {\em effectively} gives rise to DGP-type gravity
without introducing induced-gravity terms
on the brane. In the case of the sharp conical bulk $(\rho_0\ll r_c\lesssim r_0)$,
we have obtained the following result in the regime where linearization is justified:
\begin{itemize}
\item the 4D Brans-Dicke theory with $\omega_{{\rm BD}}=-r_c/r_0$ for $\mathtt{r}\ll r_c, r_0$;
\item 5D gravity for $r_c\ll\mathtt{r}\ll r_0$;
\item 6D gravity for $\mathtt{r}\gg r_c, r_0$.
\end{itemize}
(Of course, the second regime appears if $r_c\ll r_0$.)
The linear analysis is not valid below $R_*=(r_c^2R_g)^{1/3}$,
where $R_g$ is the gravitational radius of the source.
We have performed a seminonlinear analysis for $\mathtt{r}\ll R_*$ by
carefully taking into account quadratic terms in the brane bending.
Our finding is:
\begin{itemize}
\item 4D Einstein gravity is reproduced for $\mathtt{r}\ll R_*$.
\end{itemize}

The present model is very simple in that the background is given by
a locally flat spacetime.
In spite of this simplicity, the system is shown to have a rich and interesting structure.
It is easy to extend the model to more general background configurations
which are curved due to the cosmological constant and other fields living in the bulk.
The similar regularization scheme can be applied to such cases as well.
Bulk-cap asymmetry will then help to reproduce 4D gravity on the brane~\cite{Padilla, KoyamaKoyama}.

As mentioned in the Introduction, long distance modification of gravity
is often motivated by explaining the current acceleration of the Universe.
It would be interesting to construct cosmological solutions and
explore whether
the accelerating expansion is possible\footnote{
It has been known that the Lovelock term in the bulk leads to induced gravity on the
brane~\cite{Ig, Lovelock, Lovelock2}.
Self-accelerating solutions were found in the framework of 6D Gauss-Bonnet braneworlds
in~\cite{Charmousis}.
} due to gravity leakage in the present model.
We hope to address this issue in the near future.

%


\acknowledgments
I would like to thank Tetsuya~Shiromizu for a careful reading of the manuscript
and Kei-ichi~Maeda for helpful comments on the scalar-tensor theories of gravity.
I am supported by the JSPS under Contact No.~19-4199.

\appendix

\section{Linear perturbations}\label{app:per}

\subsection{Scalar perturbations}\label{app:sc}

In the longitudinal gauge ($B=E=0$), the perturbed 6D Einstein tensor
for scalar-type variables is given by
\begin{eqnarray}
\delta G_{\rho\rho}&=&\frac{4}{\rho}\Psi'+3\Box\Psi-\Box \Phi,\label{app:rhorho}
\\
\delta G_{\mu \rho}&=&\partial_{\mu}\left(
-3\Psi'+ \Phi'+\frac{1}{\rho} \Phi+\frac{1}{\rho} \Gamma
\right),\label{app:murho}
\\
\delta G_{\varphi}^{\;\varphi}&=&4\Psi''+3\Box\Psi+\Box \Gamma,\label{app:phph}
\\
\delta G_{\mu\nu}&=&(\eta_{\mu\nu}\Box-
\partial_{\mu}\partial_{\nu})\left( 2\Psi- \Phi+ \Gamma\right)
\nonumber\\&&\hspace{-6mm}
+\left(
3\Psi''- \Phi''+\frac{3}{\rho}\Psi'-\frac{2}{\rho} \Phi'-\frac{1}{\rho} \Gamma'
\right)\eta_{\mu\nu}.\label{app:munu}
\end{eqnarray}
Here we do not include the components $\delta g_{\rho\varphi}:=\rho^2 T$
and $\delta g_{\mu\varphi}:=\partial_{\mu}D$. As we will show below, these two variables
can be gauged away.

The traceless part of Eq.~(\ref{app:munu}) leads to the relation~(\ref{rel-scalar2}).
Eqs.~(\ref{rel-scalar2}) and~(\ref{app:murho}) then combine to give Eq.~(\ref{rel-scalar1}).
Finally, using Eqs.~(\ref{rel-scalar2}),~(\ref{app:rhorho}) and~(\ref{app:phph})
we eliminate $\Phi$ and $\Gamma$ to obtain Eq.~(\ref{master-scalar}).
The trace part of Eq.~(\ref{app:munu}) is derived from Eq.~(\ref{app:murho}) and hence is redundant.

The gauge transformation
$x^{\mu}\to x^{\mu}+\partial^{\mu}\delta x$ and $\rho\to \rho+\delta \rho$ yields,
at linear order in the gauge parameters,
\begin{eqnarray}
&&\Psi\to\Psi,\quad
E\to E-\delta x,\quad
B\to B-\delta \rho-\delta x',
\nonumber\\&&\qquad
\Gamma\to\Gamma-\delta \rho' ,\quad
\Phi\to\Phi+\delta \rho/\rho.
\end{eqnarray}
Note that these metric perturbations do not transform under
the gauge transformation $\varphi\to\varphi+\delta\varphi$.

The Gaussian normal gauge is defined by
$\overline{\delta g}_{\rho \rho}=\overline{\delta g}_{\rho \mu}=0$
and the brane location kept unperturbed, $\overline{\rho}=\rho_0$.
The gauge parameters giving the transformation to the Gaussian normal gauge satisfy
\begin{eqnarray}
\delta\rho|_{\rho_0} = -\zeta_\rml,\quad\delta \rho'|_{\rho_0}=\Gamma|_{\rho_0},
\nonumber\\
\delta x|_{\rho_0}=0, \quad\delta x'|_{\rho_0}=\zeta_\rml,
\end{eqnarray}
where $\zeta_\rml$ is the perturbation of the brane location in the longitudinal gauge.
We have chosen $\delta x|_{\rho_0}=0$ so that $\overline{E}=0$ on the brane.
The above equations are for the cap. The equations for the bulk are simply derived
by replacing $\rho$ with $r$.

Finally, we turn to the $(\rho\varphi)$ and $(\mu\varphi)$ components of the metric perturbations.
Under the gauge transformation $\varphi\to\varphi+\delta\varphi$, these components transform as
$T\to T-\partial_{\rho}\delta\varphi$ and $D\to D-\rho^2\delta\varphi$.
First we eliminate $T$ using the gauge degree of freedom, and then the
Einstein equations imply that $\partial_{\rho}\left(\rho^{-2}\Box D\right)=0$.
The solution is given by $D=\rho^2 D_0(x)$, but this can also be eliminated
using the residual gauge freedom $\delta\varphi=\delta\varphi(x)$.

\subsection{Vector perturbations}\label{app:vec}

The vector-type perturbations are
\begin{eqnarray}
\delta g_{\mu\nu}=2E_{(\mu,\nu)},
\quad \delta g_{\mu\varphi}=D_{\mu}, \quad\delta g_{\mu\rho} = B_{\mu},
\end{eqnarray}
where $\partial^{\mu}E_{\mu}=\partial^{\mu}D_{\mu}=\partial^{\mu}B_{\mu}=0$.
The vector gauge transformation $x^{\mu}\to x^{\mu}+\delta x^{\mu}$ leads to
\begin{eqnarray}
E_{\mu}\to E_{\mu}-\delta x_{\mu},\; D_{\mu}\to D_{\mu}, \;B_{\mu}\to B_{\mu}-
\delta x_{\mu}'.
\end{eqnarray}
The Einstein equations are
\begin{eqnarray}
(\mu\varphi):\;\;&&
D_{\mu}''-\frac{1}{\rho}D_{\mu}'+\Box D_{\mu}=0,
\\
(\mu\nu):\;\;&&
V_{\mu}'+\frac{1}{\rho}V_{\mu}=0,
\\
(\mu\rho):\;\; &&\Box V_{\mu}=0,
\end{eqnarray}
where we defined a gauge-invariant quantity $V_{\mu}:=B_{\mu}-E_{\mu}'$.
Choosing the gauge in which $B_{\mu}=0$, we obtain
\begin{eqnarray}
E_{\mu}=
\begin{cases}\displaystyle{
v_{\mu}(x)\ln(\rho/\rho_0)+u_{\mu}(x)
}
\\
\displaystyle{
v_{\mu}(x)\ln(r/r_0)+u_{\mu}(x)
}
\end{cases},
\end{eqnarray}
where the continuity across the brane has been imposed.
However, the regularity at the center of the disk requires $v_{\mu}(x)=0$ and
$u_{\mu}(x)$ can be eliminated using the residual gauge freedom $\delta x_{\mu}=\delta x_{\mu}(x)$.
The metric perturbation $D_{\mu}$ is sourced only by the $(\mu\varphi)$ component of the
brane energy momentum tensor via the junction conditions.
We ignore the corresponding component in the main text.

\subsection{Tensor perturbations}

The traceless and transverse tensor perturbation $h_{\mu\nu}$ is gauge invariant.
The equation of motion~(\ref{master-tensor}) simply follows from the linearized Einstein equations.

\section{Derivation of the junction equations~(\ref{junc_munu}) and~(\ref{junc_phiphi})}\label{app:der}

In the Gaussian normal gauge we have
\begin{eqnarray*}
\delta k_{\mu\nu}&=&\frac{1}{2}\overline{\gamma}_{\mu\nu}'-\eta_{\mu\nu}\left(
\frac{1}{2}{\overline{\gamma}_{\lambda}^{\;\lambda}}'-\overline{\Phi}'
\right)-\frac{1}{\rho_0}\overline{\gamma}_{\mu\nu},
\\
\delta k_{\varphi}^{\;\varphi}&=&-\frac{1}{2}{\overline{\gamma}_{\lambda}^{\;\lambda}}'.
\end{eqnarray*}
The gauge transformation rule~(\ref{gt}) gives
\begin{eqnarray*}
\delta k_{\mu\nu}&=&\frac{1}{2}h_{\mu\nu}'|_{\rho_0}
+{\cal D}_{\mu\nu}\zeta_\rml+\left(\overline{\Phi}'-3\Psi'|_{\rho_0}\right)
\eta_{\mu\nu}-\frac{1}{\rho_0}
\overline{\gamma}_{\mu\nu},
\\
\delta k_{\varphi}^{\;\varphi}&=&\Box\zeta_\rml-4\Psi'|_{\rho_0}.
\end{eqnarray*}
The bulk side equations can be derived similarly.
The perturbed energy-momentum tensor is given by
\begin{eqnarray*}
\delta{\cal T}_{\mu\nu}&=&\left(-\lambda-\frac{1}{2}q^{\varphi\varphi}Q^2\right)\overline{\gamma}_{\mu\nu}
-q^{\varphi\varphi}Q^2\overline{\Phi}\eta_{\mu\nu}+T_{\mu\nu},
\\
\delta{\cal T}_\varphi^{\;\varphi}&=&q^{\varphi\varphi}Q^2\overline{\Phi}+T_\varphi^{\;\varphi},
\end{eqnarray*}
where $q^{\varphi\varphi}=\rho_0^{-2}=(\beta r_0)^{-2}$.
Using the gauge transformation rule~(\ref{gt})
and the background junction conditions~(\ref{bj1}) and~(\ref{bj2}),
it is straightforward to show
\begin{eqnarray*}
&&M_\rml^4\left(\frac{1}{2}h_{\mu\nu}'|_{\rho_0}+{\cal D}_{\mu\nu}\zeta_\rml\right)
\nonumber\\&&\quad
-M_\rmr^4\left(\frac{1}{2}h_{\mu\nu}'|_{r_0}+{\cal D}_{\mu\nu}\zeta_\rmr\right)
+\Delta = T_{\mu\nu},
\end{eqnarray*}
where
\begin{eqnarray*}
\Delta &=&M_\rml^4\left.\left(-3\Psi'+\Phi'+\frac{1}{\rho_0}\Phi+\frac{1}{\rho_0}\Gamma\right)
\right|_{\rho_0}
\\&&\quad
-M_\rmr^4\left.\left(-3\Psi'+\Phi'+\frac{1}{r_0}\Phi+\frac{1}{r_0}\Gamma\right)
\right|_{r_0}.
\end{eqnarray*}
However, the perturbed Einstein equations $\delta G_{\mu\rho}=\delta G_{\mu r}=0$ implies that $\Delta=0$,
and therefore the above equation reduces to Eq.~(\ref{junc_munu}).
Similarly, one obtains Eq.~(\ref{junc_phiphi}).

\section{Yet another case: $\rho_0\sim r_0\ll r_c$}

Throughout the main text we have confined ourselves
to the case of the sharp conical bulk ($\rho_0/r_0=\beta\ll 1$).
In this Appendix we shall comment briefly on the different case with $\rho_0\sim r_0\ll r_c$,
using the point source example in Sec.~\ref{sec:pointsource}.

On scales much larger than $\rho_0$ and $r_0$,
Eq.~(\ref{kernel-G}) reduces to
\begin{eqnarray}
\hat{{\cal G}}(k)\simeq\frac{1}{(2\pi)^3}\left[k^2+\frac{1}{r_0r_c K_0(kr_0)}\right]^{-1},
\label{sub_G}
\end{eqnarray}
and hence the crossover scale $\tilde k_c^{-1}$ in this case is determined by
$\tilde k_c^{-2}\sim r_0r_c K_0(\tilde k_c r_0)$.
Thus, $h_{\mu\nu}\sim |\vec{x}|^{-1}$ for $|\vec{x}|\ll \tilde k_c^{-1}$
and $h_{\mu\nu}\sim |\vec{x}|^{-3}$ for $|\vec{x}|\gg\tilde k_c^{-1}$.

To see the behavior of $\Psi$ on the brane, we use
\begin{eqnarray}
K_0(kr_0)\hat\psi\simeq\frac{1}{(2\pi)^3}\left[k^2-\frac{2}{r_0r_c K_0(kr_0)}\right]^{-1}\frac{{\cal M}}{6M_4^2}.
\label{subcrit}
\end{eqnarray}
[This is derived from Eq.~(\ref{fourierJ}).]
Now the same issue as in~\cite{Kaloper} arises: moving to smaller scales from the 6D regime,
Eq.~(\ref{subcrit}) becomes quite large around $k\sim \tilde k_c$.
This means that the linearized theory breaks down when the 4D regime might set in.
To get more insight into the crossover from the 6D regime to the 4D one,
we must go beyond the linear analysis.
It is interesting, however, to note that for $k\gg \tilde k_c$ the first terms in
the square brackets in Eqs.~(\ref{sub_G}) and~(\ref{subcrit}) win, so that
4D Einstein gravity is likely to be recovered on the brane.


\end{document}